\documentclass[12pt,a4paper]{article}

\usepackage{epsfig}
\epsfxsize=\textwidth
\renewcommand{\includegraphics}[1]{\epsfbox{#1}}

\makeatletter
\newcounter{lineno}
\def\verbatimlisting#1{\setcounter{lineno}{0}%
    \begingroup \@verbatim \frenchspacing \@vobeyspaces \parindent=20pt
    \everypar{\stepcounter{lineno}\llap{\thelineno\ \ }}\input#1
    \endgroup
}
\makeatother

\newcommand{\del}{\varepsilon}
\newcommand{\adot}[2]{\langle #1 , #2 \rangle}
\newcommand{\reduce}{{\sc reduce}}
\newcommand{\D}[2]{{\frac{\partial #1}{\partial #2}}}

\newcommand{\DD}[2]{{\frac{\partial^2 #1}{\partial #2^2}}}

\newcommand{\Ord}[1]{{\cal O}\left(#1\right)}
\newcommand{\cD}{{\cal D}}
\newcommand{\cE}{{\cal E}}

\newcommand{\cG}{{\cal G}}
\newcommand{\cI}{{\cal I}}
\newcommand{\cJ}{{\cal J}}
\newcommand{\cL}{{\cal L}}
\newcommand{\cM}{{\cal M}}
\newcommand{\cN}{{\cal N}}

\newcommand{\tv}{\tilde{\vec v}}
\newcommand{\tg}{\tilde{\vec g}}
\newcommand{\tz}{\tilde{\vec z}}
\newcommand{\pz}{{\vec z}'}
\newcommand{\pv}{{\vec v}'}
\newcommand{\pg}{{\vec g}'}

\newcommand{\vs}{{\vec s}}
\renewcommand{\vec}[1]{\mbox{\boldmath$#1$}}

\begin{document}
\title{\sf Computer algebra derives correct initial conditions for
low-dimensional dynamical models}
\author{A.J.~Roberts\thanks{Dept.\ Mathematics \& Computing,
University of Southern Queensland, Toowoomba 4350, \textsc{Australia}.
E-mail: \texttt{aroberts@usq.edu.au}}}
\date{9 June 1997}
\maketitle

\paragraph*{PACS:} 02.70.Rw, 04.20.Ex, 02.30.Mv, 05.45.+b

\paragraph*{Keywords:} computer algebra, initial conditions,
low-dimensional modelling, centre manifold, bifurcation.

\begin{abstract}
To ease analysis and simulation we make low-dimensional models of
complicated dynamical systems.  Centre manifold theory provides a
systematic basis for the reduction of dimensionality from some
detailed dynamical prescription down to a relatively simple model.  An
initial condition for the detailed dynamics also has to be projected
onto the low-dimensional model, but has received scant attention.
Herein, based upon the reduction algorithm in~\cite{Roberts96a}, I
develop a straightforward algorithm for the computer algebra
derivation of this projection.  The method is systematic and is based
upon the geometric picture underlying centre manifold theory.  The
method is applied to examples of a pitchfork and a Hopf bifurcation.
There is a close relationship between this projection of initial
conditions and the correct projection of forcing onto a model.  I
reaffirm this connection and show how the effects of forcing, both
interior and from the boundary, should be properly included in a
dynamical model.
\end{abstract}

\tableofcontents

\section{Introduction}
\label{Sintro}

Ordinary differential equations or partial differential equations are
used to describe dynamics in the physical world.  But many modes of
behaviour are of little physical interest in particular applications.
The essential dynamical behaviour of the system is then determined by
the evolution of a subset of the possible modes; for example, the
flight of a ball is dictated solely by its velocity and spin, and
hardly at all by any internal visco-elastic vibrations.  Typically we
say that the state of the system $\vec u$ is some function,
\begin{equation}
	\vec u=\vec v(\vec s)\,,
	\label{Evs}
\end{equation}
of the few interesting modes $\vec s$.  Then a model of the dynamics,
the rigid-body dynamics of a ball for example, is governed by the
low-dimensional dynamical system
\begin{equation}
	\dot{\vec s}=\vec g(\vec s)\,,
	\label{Egs}
\end{equation}
where the overdot denotes $d/dt$.  Rapid damping typically
characterises the modes that need to be eliminated from consideration.
When many modes are heavily damped, trajectories are rapidly attracted
to some low-dimensional subset of the state space, parameterized as in
(\ref{Evs}), a so-called invariant manifold \cite[\S1.1C]{Wiggins90}.
This geometric picture of exponential collapse to a smooth invariant
manifold is at the heart of the application of centre manifolds
\cite{Carr81,Coullet83,Roberts97a} to the rational construction of
low-dimensional models~(\ref{Egs}) by the elimination of physically
uninteresting fast modes of behaviour.

The same concepts also lie behind other recent innovations in forming
low-dimensional models of dynamics: inertial manifolds by Temam
\cite{Temam90} and others; the use of centre-unstable manifolds
\cite{Armbruster89,Cheng92,Chow88}, more general invariant manifolds
\cite{Roberts89,Roberts90,Watt94b} and the so-called nonlinear
Galerkin method \cite{Marion89,Temam89,Foias88b,Luskin89,Foias94}.

In the earlier paper \cite{Roberts96a} I reported an iterative method,
based directly upon the residuals of the governing differential
equations, for the construction of such low-dimensional, dynamical
models.  The evaluation of the residuals is a routine algebraic task
which is easily programmed into computer algebra; programmed without
having to become involved in all the messy details of asymptotic
expansions.  Two examples were discussed in~\cite{Roberts96a}: a model
displaying the pitchfork bifurcation of a modified Burger's equation;
and the long-wave lubrication dynamics of a thin film of fluid.

However, it is not sufficient to just model the dynamics.  To form a
complete problem, an initial condition, say $\vec s(0)=\vec s_0$, has
to be specified for the model~(\ref{Egs}).  For very low-dimensional
models with simple attractors there is perhaps little motivation,
which may explain the lack of attention given to this issue.  However,
for models of higher dimension, initial conditions have long-lasting
effects and need to be modelled correctly.  This is seen in examples
such as: the approach to limit cycles \cite{Winfree74,Guckenheimer75};
the quasi-geostrophic approximation \cite[e.g.]{Lorenz86}; long-wave
models of fluid films \cite[e.g.]{Roy96} and of dispersion in channels
\cite[e.g.]{Mercer94a}; and the concept of ``initial slip'' in some
disciplines \cite{Grad63,Haake83,Geigenmuller83}.

In \cite[\S2]{Roberts89b} I introduced a simple dynamical system to
illustrate the correct treatment of initial conditions when forming a
low-dimensional model.  I now summarise that example.  Consider the
2-D dynamical system
\begin{eqnarray}
	\dot x & = & -xy\,,
	\label{Extdot}  \\
	\dot y & = & -y+x^2-2y^2\,.
	\label{Eytdot}
\end{eqnarray}
Linearly, $y$ decays exponentially quickly to leave $x$ as the
dominant mode appearing in the long-term evolution.  Nonlinear theory,
see Carr \cite{Carr81} for example, asserts that the long-term
evolution actually takes place on the parabolic centre manifold,
$\cM_c$, described parametrically as
\begin{equation}
	x=s\,,\quad y=s^2\,.
	\label{Ecmt}
\end{equation}
Exponentially quickly solutions of~(\ref{Extdot}--\ref{Eytdot})
approach solutions on $\cM_c$ whose evolution is described by
the \emph{low-dimensional model}
\begin{equation}
	\dot s=-s^3\,.
	\label{Estdot}
\end{equation}
The question is: given some initial state for the detailed dynamical
system~(\ref{Extdot}--\ref{Eytdot}), say $(x_0,y_0)$ not on $\cM_c$,
what initial value of $s$, say $s_0$, is appropriate to use with the
model~(\ref{Estdot}) in order to guarantee fidelity between
predictions of the model and the original?  In
\cite[\S2.1]{Roberts89b} I showed that this is achieved by setting
\begin{equation}
	s_0=x_0-x_0(y_0-x_0^2)+\Ord{\del^2}\,,
	\label{Epft}
\end{equation}
where $\del$ measures the distance from the given initial condition
to the centre manifold.  Alternatively, in a form we use in later
generality, this projection from the given $(x_0,y_0)$ to the
correct $(s_0,s_0^2)$ on $\cM_c$ is orthogonal to the vector $\vec
z(s)\approx(1,-s)$.  Only then does the predictions of the model
accurately describe the long-term evolution of the original.

How in general do we find initial conditions to use with a model to
make correct long-term forecasts?  The Relevance Theorem of centre
manifold theory \cite{Carr81,Roberts96a} assures us that there is
indeed a particular solution of the low-dimensional model on $\cM_c$
which is approached exponentially by every trajectory of the full
dynamical system.  As developed in~\cite{Roberts89b}, the geometric
picture of evolution near the centre manifold suggests a method of
analysis.  The algebra is based upon how trajectories near to the
centre manifold evolve, identifying which ones approach each other
exponentially quickly and thus have the same long-term evolution.  For
example, in the long-term dispersion down a channel
\cite{Mercer94a,Watt94b} we can discern the difference between dumping
contaminant into the slow moving flow near the bank and into the fast
core flow.  Normal form transformations \cite{Cox93b} also support
this projection of initial conditions onto the centre manifold.  In
this paper I simplify the main results of \cite[pp65--6]{Roberts89b},
\S\S\ref{SSproj}, and in \S\S\ref{SSitp} show how to solve the new
equations using an iterative scheme based on that developed in
\cite{Roberts96a} to derive the dynamical model.

However, the centre manifold based analysis reported to date has
always dealt with the issue of initial conditions for models based
upon \emph{slow} modes with near zero growth rate.  In general, a
centre manifold model will involve the oscillating dynamics of a Hopf
bifurcation or of the loss of stability to travelling waves.  In
\S\ref{Shopf} we also consider what extensions need to be made to
the analysis and the iterative algorithm to provide correct initial
conditions to models of such oscillations.

Lastly, one attribute of basic centre manifold theory is that it
applies to autonomous dynamical systems.  But the modelling of
dynamics with time-dependent forcing is of considerable interest.  In
the presence of forcing a system is pushed away from the centre
manifold as quantified by Cox \& I \cite{Cox93b}.  Thus there is a
close connection between the geometric projection of initial
conditions and the appropriate projection of forcing onto the model
\cite[\S7]{Roberts89b}.  There is also many interesting issues in the
modelling of noisy dynamical systems, as expressed by stochastic
differential equations \cite{Chao95}.  In \S\S\ref{SSforc} I reaffirm
the connection and show that a new normalisation of the initial
condition projection onto the centre manifold makes the projection of
forcing significantly simpler.  These are applied in \S\S\ref{SSifbc}
to the modified Burger's dynamics to show how both interior and
boundary forcing are treated.  These methods should apply to
interesting questions such as: what influence may turbulence have on
dispersion?  and how does substrate roughness affect the flow of thin
films?

\section{Geometric basis of initial condition projection}
\label{Sproj}

The aim of this section is to introduce some of the concepts and
details of modelling initial conditions of dynamical systems.  The
presentation improves on earlier work reported elsewhere,
predominantly in \cite{Roberts89b,Cox91,Chao95}, and is adapted to
a simple computational approach.

Consider a general autonomous dynamical system written in a form
relative to some fixed point (taken to be the origin without loss of
generality):
\begin{equation}
	\dot{\vec u}=\cL\vec u+\vec f(\vec u,\epsilon)\,,
	\label{Eds}
\end{equation}
where $\vec u(t)$ is the state vector, which may be finite or infinite
dimensional, $\cL$ is the linear operator, and $\vec f$ denotes all
the nonlinear terms in the dynamical prescription with possible small
parameters $\epsilon$.  We suppose that $\cL$ has $m$ eigenvalues with
zero real-part and the remaining eigenvalues have strictly negative
real-part.  The linear dynamics of $\dot{\vec u}=\cL\vec u$ then
collapse exponentially quickly onto the centre subspace $\cE_c$
spanned by the eigenvectors and possibly generalised eigenvectors of
$\cL$, namely $\vec e_j^0$.  Nonlinear theory \cite{Carr81}
correspondingly asserts that the exists an $m$-dimensional,
exponentially attractive centre manifold $\cM_c$ which has $\cE_c$ as
its tangent at the origin.  The centre manifold is parameterized by
any convenient set of $m$ parameters, say $\vs $.  Thus the
low-dimensional model of~(\ref{Eds}) is that
\begin{equation}
	\vec u=\vec v(\vs ,\epsilon)\,,
	\quad\mbox{such that}\quad
	\dot{\vs }=\cG\vs +\vec g(\vs ,\epsilon)\,,
	\label{Emod}
\end{equation}
where $\vec v$ describes the shape of $\cM_c$, $\cG$ is a linear
operator with eigenvalues all of real-part zero, and $\vec g$ is
strictly nonlinear.  Theory also asserts this model is valid
exponentially quickly.

In the earlier paper~\cite{Roberts96a} I derived a robust and
straightforward iterative algorithm to construct approximations to the
model functions $\vec v$ and $\vec g$.  I take this work as read.  We
now move on to develop the correct projection of initial conditions
for~(\ref{Eds}), and then how to project a forcing superimposed on the
autonomous dynamical system.

For simplicity I first assume that the critical eigenvalues of
$\cL$ are all zero.  The case of non-zero, pure imaginary eigenvalues
has extra complicating details and is addressed in \S\ref{Shopf} where
we investigate the specific example of a Hopf bifurcation.

\subsection{Evolution near the centre manifold}
\label{SSproj}

As in \cite[\S5.1]{Roberts89b}, consider the evolution of a point on
the centre manifold, $\cM_c$, and the evolution of any neighbouring
point.  Let $\vec n$ denote the small vector joining these points,
then, under the flow of the dynamics of~(\ref{Eds}), $\vec n$
satisfies the linear equation
\begin{equation}
	\frac{d\vec n}{dt}=\cJ\vec n\,,
	\quad\mbox{where}\quad
	\cJ=\cL+\cN\,,
	\label{Ejac}
\end{equation}
is the Jacobian of~(\ref{Eds}) evaluated on $\cM_c$ and where
$\cN=\partial\vec f/\partial\vec u$ is the Jacobian of the nonlinear
and parameter dependent terms.  Note that $\cN$ and hence $\cJ$ are
functions of position $\vs $ on the centre manifold.  It is
useful to imagine any given $\vec n$ to be a function of position on
$\cM_c$ rather than time; in this case we deduce by the chain rule
that
\begin{displaymath}
	\frac{d\cdot }{dt}=\D{\cdot }{\vs }\left(\cG\vs +\vec g\right)
	=\left(\cG_{jk}s_k+g_j\right)\D{\ }{s_j}\,,
\end{displaymath}
using the summation convention hereafter.  The projection of initial
conditions onto $\cM_c$ is done along what have been termed
``isochronic manifolds'', denoted herein by $\cI$.  Such a projection
of initial conditions is also supported by normal form transformations
as discussed Cox \& Roberts~\cite{Cox93b}.  The most well known
isochronic manifold is simply the stable manifold of the origin,
$\cM_s$: since the origin is an equilibrium in the model, that
trajectories starting on $\cM_s$ exponentially quickly approach the
origin is precisely the requirement for the isochronic manifold of the
origin's fixed dynamics.

Under the evolution in the neighbourhood of $\cM_c$ the displacement
vector $\vec n$ will do either of two things: the tip of $\vec n$ off
the manifold will approach $\cM_c$ exponentially quickly but in
general it will have slipped from the base point on $\cM_c$ and thus
$\vec n$ will exponentially quickly become tangent to $\cM_c$;
alternatively, if $\vec n$ is aligned just right then the tip will
approach the base while remaining transverse to $\cM_c$.  It is this
latter case that is of interest as an initial condition of the full
dynamics at the tip of $\vec n$ will result in a long-term evolution
that is indistinguishable from that of the evolution of the base
point.  Thus the base point forms the appropriate initial condition
for the low-dimensional dynamics on $\cM_c$.  The set of all such tip
points that exponentially approach $\cM_c$, while $\vec n$ remains
transverse, forms the isochronic manifold $\cI$ for any specified
initial condition on $\cM_c$.

To describe the projection of given initial conditions $\vec u_0$ of
the full dynamics onto $\cM_c$, we use the normal vectors to the
isochronic manifolds.  More specifically we use the normal vectors of
$\cI$ at $\cM_c$.  Let $\vec n_\alpha=\vec u_0-\vec v(\vec
s_0,\epsilon)$ be a family of small vectors which span the tangent
space of $\cI$.  Then in terms of an inner product $\adot{\ }{\ }$, we
seek $m$ linearly independent vectors, say $\vec r_i$ as a function of
position $\vs $, such that
\begin{equation}
	\adot{\vec r_i}{\vec n_\alpha}
	=0\,.
	\label{Eproj}
\end{equation}
It is much easier to find the $m$ vectors $\vec r_i$ than the possible
infinity of vectors $\vec n_\alpha$.  Taking $d/dt$ of~(\ref{Eproj})
and using~(\ref{Ejac}) we deduce
\begin{equation}
	\cD\vec r_i=\vec 0
	\quad\mbox{where}\quad
	\cD=\frac{d\ }{dt}+\cJ^\dag\,,
	\label{Eadj}
\end{equation}
and $\cJ^\dag$ denotes the adjoint in the specified inner product.
This equation describes how the normal vectors $\vec r_i$ vary over
$\cM_c$.  Call $\cD$ the \emph{dual}.

We need to solve~(\ref{Eadj}).  But in general $\vec r_i$ will
inconveniently vary quickly in magnitude and possibly direction over
$\cM_c$.  Whereas all we are actually interested in is the space
spanned by $\vec r_i$, namely the tangent space to $\cI$ at $\cM_c$.
Instead we seek an equivalent basis for $\cI$, one which varies
relatively slowly over $\cM_c$, by the invertible transformation $\vec
r_i=Q_{ij}\vec z_j$ for some basis vectors $\vec z_j$ also a function
of position on $\cM_c$.  It is from here that we depart significantly
from the analysis reported in my earlier work \cite[\S5]{Roberts89b}.
For a reason that becomes apparent in the next subsection, we seek the
particular basis such that
\begin{equation}
	\adot{\vec z_i}{\vec e_j}=\delta_{ij}
	\quad\mbox{where}\quad
	\vec e_j(\vs ,\epsilon)=\D{\vec v}{s_j}
	\label{Ewdef}
\end{equation}
are local tangent vectors to $\cM_c$ based upon the parameterization
of $\cM_c$; typically $\vec e_j\to\vec e_j^0$ as
$(\vs ,\epsilon)\to\vec 0$.  Substituting $\vec r_i=Q_{ij}\vec z_j$
into~(\ref{Eadj}) then leads to
\begin{eqnarray*}
	 &  & \frac{dQ_{ij}}{dt}\vec z_j+Q_{ij}\cD\vec z_j=\vec 0  \\
	 & \Rightarrow & \frac{dQ_{ik}}{dt}
	 +Q_{ij}\adot{\cD\vec z_j}{\vec e_k}=0
	 \quad\mbox{upon taking $\adot{\ }{\vec e_k}$}  \\
	 & \Rightarrow & -Q_{ij}\adot{\cD\vec z_j}{\vec e_k}\vec z_k
	 +Q_{ij}\cD\vec z_j=\vec 0
	 \quad\mbox{putting $dQ_{ij}/dt$ in the first.}
\end{eqnarray*}
But $Q_{ij}$ is an invertible matrix and thus we must solve
\begin{equation}
	\cD\vec z_j-\adot{\cD\vec z_j}{\vec e_k}\vec z_k
	=\vec 0\,,
\label{Ezeq}
\end{equation}
in conjunction with the orthonormality condition~(\ref{Ewdef}) in
order to find the basis vectors $\vec z_j$ for the isochronic
manifolds $\cI$.

Equation~(\ref{Ezeq}) has a reasonable interpretation.  The second
term just projects the residual of the dual~(\ref{Eadj})
onto $\cI$, and hence~(\ref{Ezeq}) requires all other components of
the residual to be zero.  Thus~(\ref{Ezeq}) requires that the basis
indeed twists with $\cI$, but places no restraint on how the basis
spans $\cI$; the orthonormality condition~(\ref{Ewdef}) closes the
problem to give a unique solution.

Once the basis vectors $\vec z_j$ are found, we then solve
\begin{equation}
	\adot{\vec z_j(\vs _0,\epsilon)}{\vec u_0-\vec v(\vs _0,\epsilon)}=0
	\quad\mbox{for all $j$}
	\label{Esolv}
\end{equation}
to determine the projection of a given initial state $\vec u_0$
onto an initial state $\vs _0$ for the model~(\ref{Emod}).  This
projection is linear in distance away from the centre manifold
$\cM_c$.  There will be errors quadratic in the distance.  However, in
many applications the stable manifold $\cM_s$ is precisely the linear
stable subspace $\cE_s$; hence at least near the origin we may expect
that a linear projection onto $\cM_c$ will be quite good.

\subsection{Model forcing by the same projection}
\label{SSforc}

In this subsection we consider the dynamical system~(\ref{Eds}) with a
small, of $\Ord{\del}$, forcing superimposed.  Namely we analyse
briefly
\begin{equation}
	\dot{\vec u}=\cL\vec u+\vec f(\vec u,\epsilon)
	+\del\vec p(\vec u,t,\epsilon)\,,
	\label{Efds}
\end{equation}
for small forcing $\del \vec p$.  The forcing could be deterministic,
as investigated by Cox \& I \cite{Roberts89b,Cox91}, or stochastic as
examined by Chao \& I \cite{Chao95}.  Our aim is to transform the
forcing of the detailed system~(\ref{Efds}) into a corresponding
forcing of the model~(\ref{Emod}).  That is, we seek the forced centre
manifold and the evolution thereon in the form
\begin{eqnarray}
	\vec u&=&\vec v(\vs ,\epsilon)+\del\vec w(\vs ,t,\epsilon)
	+\Ord{\del^2}\,,
	\nonumber\\
	\mbox{s.t.}\quad
	\dot{\vs }&=&\cG\vs +\vec g(\vs ,\epsilon)
	+\del\vec q(\vs ,t,\epsilon)+\Ord{\del^2}\,,
	\label{Efmod}
\end{eqnarray}
where $\vec w$ describes the displacement of $\cM_c$ and $\vec q$ is
our main interest as it describes the correct forcing to be used in
the model.

That the projection of a forcing onto a model is nontrivial was shown
in \cite[\S7.1]{Roberts89b}.  There a steady forcing of $-\del$ in
the $y$ equation~(\ref{Eytdot}) of the simple
system~(\ref{Extdot}--\ref{Eytdot}) results in the model
\begin{displaymath}
	\dot s=-s^3+\del s\,,
\end{displaymath}
which exhibits the destabilization of the origin in favour of either of
two fixed points at $s=\pm\sqrt\del$.  This result is remarkable in
that the response is comparatively large, of $\Ord{\sqrt\del}$, when
the original forcing is normal to $\cE_c$ and so usually would be
neglected by heuristic arguments.

In \cite[\S7]{Roberts89b} I argued that the projection of initial
conditions could be used to deduce how to model forcing.  This close
connection between the two processes was supported by further work by
Cox \& I \cite{Cox91}.  Here I briefly reaffirm the connection and
show why the orthonormality condition~(\ref{Ewdef}) is desirable.

To find an equation for $\vec q$, simply substitute~(\ref{Efmod})
into the original system~(\ref{Efds}) and group all terms linear in
$\del$ to deduce
\begin{equation}
	\frac{d\vec w}{dt}+\cE \vec q=\cJ\vec w+\vec p\,,
	\label{Ewqeq}
\end{equation}
where $\cE=\left[\vec e_j\right]$ is the matrix of tangent vectors, and here
\begin{displaymath}
		\frac{d\cdot }{dt}=\D{\cdot}t
		+\D{\cdot }{\vs }\left(\cG\vs +\vec g\right)\,,
\end{displaymath}
due to the direct dependency upon time introduced by the forcing $\vec
p(\vec u,t)$.  Taking $\adot{\vec z_i}{\ }$ of this equation and using
the adjoint properties we must have
\begin{equation}
	\frac{d\ }{dt}\adot{\vec z_i}{\vec w}
	-\adot{\frac{d\vec z_i}{dt}}{\vec w}
	+\adot{\vec z_i}{\vec e_j} q_j
	=\adot{ \cJ^\dag\vec z_i}{\vec w}
	+\adot{\vec z_i}{\vec p}\,.
	\label{Eawqeq}
\end{equation}
Using the orthonormality~(\ref{Ewdef}), $\adot{\vec z_i}{\vec e_j}
q_j=q_i$.  Without loss of generality, choose the
parameterization of positions near $\cM_c$ so that
\begin{equation}
	\adot{\vec z_i}{\vec w}=0\,.
	\label{Ezw}
\end{equation}
Indeed, Cox \& I \cite{Cox91} showed that this is the \emph{only}
choice for $\vec w$ that removes clumsy history dependent integrals
from the forcing $\vec q$ of the model.  Thus~(\ref{Eawqeq}) becomes
\begin{displaymath}
	q_i=\adot{\vec z_i}{\vec p}
	+\adot{\frac{d\vec z_i}{dt}+\cJ^\dag\vec z_i}{\vec w}\,.
\end{displaymath}
The last term involves $\cD\vec z_i$ which, by the projected
dual~(\ref{Ezeq}), must lie in the space spanned by the $\vec z_k$'s,
is thus orthogonal to $\vec w$, and so the last term vanishes.  Hence
the appropriate linear approximation to the forcing of the model is
simply the projection
\begin{equation}
	q_i=\adot{\vec z_i}{\vec p}\,,
	\label{Efq}
\end{equation}
for any given forcing $\vec p$ in terms of the vectors $\vec z_i$
determined for the projection of initial conditions.

\section{Initial conditions---a pitchfork bifurcation}
\label{Spit}

In \cite{Roberts96a} I described a simple and robust iterative scheme
for the computer algebra derivation of the dynamical
model~(\ref{Emod}) from the detailed system~(\ref{Eds}).  In this
section I describe how to extend the iterative scheme to derive the
projection of initial conditions.  This scheme is also eminently
suitable for computer algebra and I illustrate its application by
using the iteration to determine initial conditions for the relatively
simple pitchfork bifurcation dynamics in a specific infinite
dimensional dynamical system.  Here we consider the simpler case of
centre manifold models formed when the critical eigenvalues are
precisely zero, rather than the more complicated case of non-zero
imaginary part considered in the next section.

A summary of the iteration for the centre manifold model is as
follows.  Suppose we know an approximation to $\cM_c$ and the
evolution thereon, namely
\begin{displaymath}
	\vec u\approx\tv(\vs ,\epsilon)
	\quad\mbox{s.t.}\quad
	\dot{\vs }\approx \cG\vs +\tg(\vs ,\epsilon)\,.
\end{displaymath}
For example, usually we start the iteration with the linear
approximation: $\tv=\cE^0\vs$ and $\tg=\vec 0$.  Then we seek an
improved description, that
\begin{displaymath}
	\vec u\approx\tv+\pv
	\quad\mbox{s.t.}\quad
	\dot{\vs }\approx \cG\vs +\tg+\pg\,,
\end{displaymath}
where primes indicate small correction terms to be determined.
Substituting into the governing differential equation~(\ref{Eds}),
neglecting products of small quantities, and approximating
coefficients of primed quantities by their zeroth order
approximation, we deduce that the corrections satisfy
\begin{displaymath}
	\D\tv\vs(\cG\vs+\tg) +\cE^0\pg +\D\pv\vs\cG\vs
   =\cL\tv+\cL\pv+\vec f(\tv,\epsilon)\,.
\end{displaymath}
It is not obvious, but provided the definition of amplitudes are
arranged so that $\cG$ is in Jordan form, we may significantly
simplify the algorithm by also neglecting the term $\D\pv\vs\cG\vs$.
(It is often physically appealing to use the Jordan form because, for
example, the two amplitudes involved often represent the ``position''
and ``momentum'' of a specific mode.)  Thus, rearranging and
recognising that
\begin{displaymath}
	\D\tv\vs(\cG\vs+\tg)=\frac{d\tv}{dt}
\end{displaymath}
by the chain rule for the current approximation, we solve
\begin{equation}
	\cL\pv-\cE^0\pg=\frac{d\tv}{dt}-\cL\tv-\vec f(\tv,\epsilon)\,.
	\label{Eitgen}
\end{equation}
Recognise that the right-hand side is simply the residual of the
governing equation~(\ref{Eds}) evaluated at the current approximation.
Thus at any iteration we just deal with physically meaningful
expressions; all the complicated rearrangements of asymptotic
expansions as needed by earlier methods are absent.  All the messy
algebra in the repeated evaluation of the residuals may be left to the
computer to perform---such mindless repetition is ideal for a
computer---whereas all a human need concern themselves with is setting
up the typical solution of
\begin{displaymath}
	\cL\pv-\cE^0\pg=\mbox{residual}\,,
\end{displaymath}
and not at all with the
detailed algebraic machinations of asymptotic expansions.

Consider the following variation to Burger's equation featuring
growth, $(1+\epsilon)u$, nonlinearity, $uu_x$, and dissipation,
$u_{xx}$:
\begin{equation}
	\D ut=(1+\epsilon)u+u\D ux+\DD ux\,,
	\quad u(0,t)=u(\pi,t)=0\,,
\label{Eburg}
\end{equation}
for some function $u(x,t)$.  View this as an infinite dimensional dynamical
system, the state space being the set of all functions $u(x)$ on
$[0,\pi]$.  The above iteration scheme may be employed to find the centre
manifold dynamics near the bifurcation that takes place as $\epsilon$
crosses zero.  This application is described in detail
in~\cite{Roberts96a}: lines~1--29 of the \reduce{}\footnote{At
the time of writing, information about {\reduce} was available from Anthony
C.~Hearn, RAND, Santa Monica, CA~90407-2138, USA.  E-mail: \tt
reduce@rand.org} computer algebra program listed in \S\S\ref{SSeg}
tell us that the centre manifold is
\begin{equation}
	u=a\sin(x)+\frac{1-\epsilon/3}{6}a^2\sin(2x)
	+\frac{1-7\epsilon/12}{32}a^3\sin(3x)
	+\Ord{\epsilon^2,a^4}\,,
	\label{Ebu}
\end{equation}
when parameterized by $a$, the amplitude of the $\sin(x)$ component of
$u$.  Upon this centre manifold the low-dimensional model is that
\begin{equation}
	\dot a=\epsilon a-\frac{1-\epsilon/3}{12}a^3 +\Ord{\epsilon^2,a^4}\,.
	\label{Ebg}
\end{equation}
This model, for example, predicts the pitchfork bifurcation as
$\epsilon$ crosses zero.

\subsection{Iterative algorithm for the projection}
\label{SSitp}

Having outlined the iteration scheme to find the centre manifold
model, we now turn to implementing a similar but novel iteration
scheme to determine the vectors $\vec z_j$ that govern the projection
of initial conditions.  The scheme is illustrated by applying it to
the pitchfork bifurcation in~(\ref{Eburg}).

We need to solve~(\ref{Ezeq}) subject to the orthonormality
condition~(\ref{Ewdef}).  The method of solution is to iteratively
improve an approximation based upon the residuals of the equations.
We start the iteration with the linear approximation that $\vec z_j$
are eigenvectors or generalised eigenvectors, $\vec z_j^0$, of
$\cL^\dag$, the adjoint linear operator.  Suppose that at some later
stage we know an approximation $\vec z_j\approx\tz_j$, we then seek an
improved approximation
\begin{equation}
	\vec z_j\approx\tz_j+\pz_j\,.
	\label{Ezap}
\end{equation}
Firstly, substituting into the orthonormality condition~(\ref{Ewdef}) gives
\begin{displaymath}
	\adot{\pz_i}{\vec e_j}=\delta_{ij}-\adot{\tz_i}{\vec e_j}\,.
\end{displaymath}
Approximating the coefficient of the primed correction quantity then
shows that we impose on $\pz_i$ the requirement that
\begin{equation}
	\adot{\pz_i}{\vec e_j^0}=\delta_{ij}-\adot{\tz_i}{\vec e_j}\,.
	\label{Epz}
\end{equation}
Secondly, substituting~(\ref{Ezap}) into the projected dual
equation~(\ref{Ezeq}) and dropping products of correction terms gives
\begin{equation}
	-\cD\pz_j+\adot{\cD\tz_j}{\vec e_k}\pz_k
	+\adot{\cD\pz_j}{\vec e_k}\tz_k
	= \cD\tz_j-\adot{\cD\tz_j}{\vec e_k}\tz_k\,.
	\label{Edzl}
\end{equation}
Approximating all coefficients of primed quantities, all appearing on
the left-hand side, by their zeroth-order approximation, we seek a
correction such that
\begin{displaymath}
	-\cL^\dag\pz_j+\adot{\cL^\dag\vec z_j^0}{\vec e_k^0}\pz_k
	+\adot{\cL^\dag\pz_j}{\vec e_k^0}\vec z_k^0
	= \cD\tz_j-\adot{\cD\tz_j}{\vec e_k}\tz_k\,.
\end{displaymath}
The two inner products on the left-hand side vanish as they both may
be transformed to a form $\adot{\ }{\cL\vec e_k^0}$ which is zero as
$\vec e_k^0$ is a critical eigenvector of $\cL$.  Thus we solve the
linear equation
\begin{equation}
    -\cL^\dag\pz_j= \cD\tz_j-\adot{\cD\tz_j}{\vec e_k}\tz_k\,,
	\label{Epzeq}
\end{equation}
for the corrections $\pz_j$.  Thus the corrections are simply driven
by the residual of the projected dual~(\ref{Ezeq}) evaluated at the
current approximation as appears on the right-hand side
of~(\ref{Epzeq}).  One uncomfortable feature of~(\ref{Epzeq}) is that
sometimes during the course of the iteration there is a component of
$\vec z_k^0$ in the right-hand side---it should be ignored and
ultimately it will vanish as the iteration proceeds.

\subsection{The example pitchfork bifurcation}
\label{SSeg}

A principal reason for adopting this approach is because the iteration
is simply implemented in computer algebra.  I discuss the
implementation of the iterative algorithm when applied to determining
initial conditions for the model~(\ref{Ebg}) of the modified Burger's
equation~(\ref{Eburg}).

Based upon the above derivation, the general outline of the algorithm
is:
\begin{enumerate}\sf

\item find the centre manifold and the
evolution thereon;

\item  initialisation and linear approximation;

\item  {\tt repeat until} residuals are small enough;
\begin{enumerate}
	\item  compute normality and adjoint residuals,

	\item  compute projected adjoint residual,

	\item  solve for the correction and update approximation.

\end{enumerate}
\end{enumerate}
In practise, the iteration for the initial condition projection could
be intertwined with the iteration for the centre manifold model.
However, here we keep them separate for clarity.

Implemented in \reduce{} for Burger's equation~(\ref{Eburg}) the
algorithm may look like the second part of the following.
{\footnotesize \verbatimlisting{biwc.red} }
Observe the how lines~32--44 of this program
implements the algorithm for the initial condition projection.
\begin{enumerate}
\item $\ell$4--29 find the centre manifold and the evolution thereon
using the iterative algorithm of~\cite{Roberts96a} as outlined at the
start of \S\ref{Spit};

\item $\ell$33--38 Initialisation.

\begin{itemize}

\item $\ell$33, the local tangent vector to $\cM_c$, namely
\begin{displaymath}
	e(x)=\sin x +\frac{1}{3}a\sin(2x) +\frac{3}{32}a^2
	+\Ord{a^3+\epsilon^{3/2}}\,,
\end{displaymath}
\item $\ell$35--36 defines \verb|lainv|, the inverse of the adjoint
operator $\cL^\dag$.  Here, under the obvious inner product
\begin{displaymath}
	\adot uv=\frac{2}{\pi}\int_0^\pi uv\,dx\,,
\end{displaymath}
$\cL$ is self adjoint so \verb|lainv| is identical to \verb|linv|,
except that we need to neglect any component in $z^0(x)=e^0(x)=\sin x$ as
commented upon earlier.

\item  $\ell$38 gives the initial linear approximation to the projection
\begin{displaymath}
	z\approx z^0=\sin x\,.
\end{displaymath}
 \end{itemize}

	\item  $\ell$39--44 perform the iteration until the residuals
	are negligible according to the \verb|let| statement of $\ell$23;
	\begin{enumerate}
		\item $\ell$40--41 compute normality~(\ref{Ewdef}) and
		dual~(\ref{Eadj}) residuals for the current approximation
		direct from their equations,

		\item $\ell$42 computes the projected dual
		residual~(\ref{Ezeq}),

		\item  $\ell$43 solves for the correction and updates the
approximation
		to the projection vector $z(x)$.

	\end{enumerate}
\end{enumerate}
Running this program shows that
\begin{eqnarray}
	z&=&\left(1+\frac{a^2}{18}\right)\sin x
	-\frac{1+\epsilon/3}{6}a\sin(2x)
	+\frac{1+5\epsilon/4}{96}a^2\sin(3x)
	\nonumber\\&&
	+\Ord{a^3,\epsilon^2}\,.
	\label{Ezeg}
\end{eqnarray}
The task of finding the correct initial condition for the
model~(\ref{Ebg}) is then the following.  Given an initial condition
for the Burger's equation~(\ref{Eburg}), namely that $u=u_0(x)$ at
$t=0$, we project onto the centre manifold by solving for the
amplitude $a_0$ in the nonlinear equation
\begin{equation}
	\adot{z(a_0,x)}{u_0(x)-v(a_0,x)}=0\,.
	\label{Eaeq}
\end{equation}
An iterative approach will usually suffice to solve this.  Starting
with the approximation $\tilde a_0=\adot{z^0}{u_0}$, successive
corrections may be computed as
\begin{displaymath}
	a'_0=\adot{z(\tilde a_0,x)}{u_0(x)-v(\tilde a_0,x)}\,.
\end{displaymath}
For example, if $u_0=\alpha\sin x$ for some particular $\alpha$, then
\begin{equation}
	a_0=\alpha+\frac{1}{36}\alpha^3+\Ord{\alpha^4,\epsilon^2}\,,
	\label{Ealph}
\end{equation}
and \emph{not} simply $a_0=\alpha$ as would be implied by a
direct application of the definition of amplitude $a$.

Of course in this particular application the issue of the precisely
correct initial condition is of little interest because the ultimate
fate of the dynamics is absorption by a stable fixed point and an
incorrect initial condition just causes a small error in the timing of
the approach.  However, in more complicated dynamical models with
non-trivial long-term dynamics, for example in chaotic models or in
the shear dispersion of contaminant in a pipe or channel, errors in
the initial condition can cause significant long-term errors in the
predictions of a model.  But before moving on to the analysis of such
a problem, we investigate in the next subsection the projection of
forcing in this modified Burger's equation~(\ref{Eburg}).

\subsection{Forcing in the interior and on boundaries}
\label{SSifbc}

The forced equation we consider briefly in this subsection is~(\ref{Eburg})
with some unspecified small forcing $\del p(u,x,t)$, namely
\begin{equation}
	\D ut=(1+\epsilon)u+u\D ux+\DD ux+\del p(u,x,t)\,.
\label{Efburg}
\end{equation}
Then by the arguments of \S\ref{SSforc} the forcing of the
model~(\ref{Ebg}) turns it into
\begin{equation}
	\dot a=\epsilon a-\frac{1-\epsilon/3}{12}a^3
	+\adot{z}{\del p(v,x,t)}
	+\Ord{\epsilon^2,a^4,\del^2}\,.
	\label{Efbg}
\end{equation}
For example, a spatially uniform, additive forcing $\del p(t)$ induces
a multiplicative forcing as in
\begin{displaymath}
	\dot a=\epsilon a-\frac{1-\epsilon/3}{12}a^3
	+\left(\frac{4}{\pi}+\frac{17+5\epsilon/4}{72\pi}a^2\right)\del p
	+\Ord{\epsilon^2,a^4,\del^2}\,.
\end{displaymath}
Also, as in the example near the start of \S\ref{SSforc}, here a
forcing proportional to $\sin(2x)$ may destabilize the origin.

The above results on forcing in the interior of the domain are
straightforward given the analysis of \S\ref{SSforc}.  A little more
subtle is the effects of forcing in the boundary conditions.  For
example, (\ref{Efburg}) may have forced boundary conditions such as
\begin{equation}
	u(0,t)=\del p_0(t)\,,\quad
	u(\pi,t)=\del p_\pi(t)\,.
	\label{Efbbc}
\end{equation}
For the moment, assume there is no forcing in the interior---these are
the only forcing terms.

One heuristic approach is to turn these boundary conditions into
interior Dirac delta function forcing of the same problem but with
homogeneous boundary conditions.  In essence, this approach forces an
extremely thin boundary layer at $x=0^+$ between the boundary value of
$u(0)=0$ and an interior value of $u(0^{++})=\del p_0$, and similarly
near $x=\pi$.  We try the forcing $\del
p=A\delta'(x-0^+)+B\delta'(x-\pi^-)$ in~(\ref{Efburg}) with the
homogeneous boundary conditions of~(\ref{Eburg}).  Here $\delta'(x)$
denotes the derivative of the Dirac delta function.  Then integrating
$x$ times (\ref{Efburg}) over the extremely small interval
$[0,0^{++}]$ leads to $u(0^{++})=-A$ as the effective boundary value
of $u$ at $x=0$.  Similar integration near $x=\pi$ leads to
$u(\pi^{--})=B$ as an effective boundary value.  Thus to match the
forced boundary conditions~(\ref{Efbbc}), we choose
\begin{equation}
	\del p(x,t)=-\del p_0(t)\delta'(x-0^+)
	+\del p_\pi(t)\delta'(x-\pi^-)\,.
	\label{Edelf}
\end{equation}
Then the projection of such an ``interior'' forcing onto the model,
equation~(\ref{Efbg}), leads to
\begin{eqnarray}
	\dot a&=&\epsilon a-\frac{1-\epsilon/3}{12}a^3
	+\frac{2}{\pi}\left(1+\frac{25}{288}a^2\right)\del (p_\pi+p_0)
	\nonumber\\&&
	+\frac{1+\epsilon/3}{3\pi}a\del (p_\pi-p_0)
	+\Ord{\epsilon^2,a^3,\del^2}\,.
	\label{Eblfg}
\end{eqnarray}
Notice that an asymmetry in the boundary forcing, $p_\pi\neq p_0$, may
destabilize the fixed point at the origin.

Another more systematic approach identifies where inhomogeneous
boundary conditions such as~(\ref{Efbbc}) affect the earlier analysis.
We now do this.  Realise that with forcing in the boundaries the
differential Jacobian term $\cJ w$ in~(\ref{Ewqeq}) then comes with
the attached boundary conditions that $w(0,t)=p_0$ and
$w(\pi,t)=p_\pi$.  Consequently, here the integration by parts in
going from~(\ref{Ewqeq}) to~(\ref{Eawqeq}) introduces extra terms as
\begin{displaymath}
	\adot{z}{\cJ w}=\adot{\cJ^\dag
	z}{w}-\frac{2}{\pi}\left[z_xw\right]_0^\pi\,.
\end{displaymath}
Hence the forcing of the model is not just the projection of the
interior forcing, $\adot zp$, but instead contains extra terms:
\begin{equation}
	q=\adot zp +\frac{2}{\pi}\left[z_x(0)p_0(t)- z_x(\pi)p_\pi(t)\right]\,.
	\label{Eibcf}
\end{equation}
After substituting in the expression~(\ref{Ezeg}) for $z$, this agrees
precisely with~(\ref{Eblfg}).

\section{Correct phase near a Hopf bifurcation}
\label{Shopf}

In the analysis and example of the previous sections, the dynamics on
the centre manifold have been based on the critical eigenvalues being
precisely 0, not the more general case where just the real part of the
eigenvalues are zero but the imaginary part is non-zero.  However,
because an eigenvalue with a non-zero imaginary part is always
associated with oscillations, such a case is important in practise as
it arises in the common transition from steady to oscillatory
dynamics.  Because it is significantly more difficult to determine the
projection of initial conditions for such oscillatory dynamics, I
describe a specific implementation in this section.  Note that in a
Hopf bifurcation, as recognised by Winfree~\cite{Winfree74} and
Guckenheimer~\cite{Guckenheimer75}, unless a good initial condition is
found the phase between the model and the actual dynamics are
irretrievably different.

As an example, consider the dynamics of the following dynamical
system
\begin{equation}
	\dot{\vec u}=\left[
	\begin{array}{ccc}
		-1 & -1 & 0  \\
		2 & 1 & 0  \\
		1 & 2 & -1
	\end{array}
	\right]\vec u +\left[
	\begin{array}{c}
		\epsilon u_1 -2u_1u_3  \\
		2u_1u_3  \\
		u_2^2
	\end{array}
	\right]\,,
	\label{Ehopfewig}
\end{equation}
where $\epsilon$ is a control parameter.  It is straightforward to
discover that when $\epsilon$ is zero, the critical value, the linear
operator has eigenvalues $\pm i$ and $-1$.  Thus there exists a centre
manifold corresponding to the two eigenvalues $\pm i$.  The mode with
eigenvalue $-1$ is representative of the many exponentially decaying
modes we find in real applications.

\subsection{The example centre manifold model}
\label{Shocm}

Our interest herein lies in the provision of correct initial
conditions for a low-dimensional model, not immediately in the
construction of the model.  Hence, in this subsection I record the
principal features of the centre manifold model of~(\ref{Ehopfewig})
and do not describe its derivation.

The eigenvectors corresponding to the critical eigenvalues enables us
to construct the centre eigenspace, $\cE_c$, the linear approximation
to the centre manifold.  The eigenvectors of $\lambda=\pm i$ are
$(2,-2\pm 2i,-3\pm i)$ and we use the real and imaginary parts of
these eigenvectors to span $\cE_c$.  A linear approximation to the
centre manifold is then
\begin{displaymath}
	\cE_c=\left\{x\vec e_x^0 +y\vec e_y^0 \mid \forall x,y\right\}
	\quad\mbox{where}\quad
	\vec e_x^0=\left[
	\begin{array}{c}
		2  \\
		-2  \\
		-3
	\end{array}
	\right]\,,\quad
	\vec e_y^0=\left[
	\begin{array}{c}
		0  \\
		2  \\
		1
	\end{array}
	\right]\,.
\end{displaymath}
Then in terms of $\vec s=(x,y)$, the linear dynamics on the centre
manifold are the oscillations described by
\begin{equation}
	\dot{\vec s}=\cG\vec s\,,
	\quad\mbox{where}\quad
	\cG=\left[
	\begin{array}{cc}
		0 & -1  \\
		1 & 0
	\end{array}
	\right]\,.
	\label{Eholind}
\end{equation}
In a general problem the centre subspace is described by $\vec u=\vec
e_j^0s_j$.  Hence the linear evolution equation $d\vec u/dt=\cL\vec u$
becomes
\begin{displaymath}
	\cL\vec e_k^0s_k=\frac{d\vec u}{dt}
	=\vec e_j^0\frac{ds_j}{dt}
	\approx \vec e_j^0\cG_{jk}s_k\,.
\end{displaymath}
Since this holds for all sufficiently small $s_k$, we deduce the
generalised basis vectors for the centre subspace satisfy
\begin{equation}
	\cL\vec e_k^0=\vec e_\ell^0\cG_{\ell k}\,.
	\label{Egbas}
\end{equation}
Similarly look at the linear equation for the leading order projection
vectors $\vec z_j^0$.  At leading order the projected
dual~(\ref{Ezeq}) becomes
\begin{displaymath}
	\cL^\dag\vec z_j^0-\adot{\cL^\dag\vec z_j^0}{\vec e_k^0}\vec z_k^0
	=0\,,
\end{displaymath}
after dropping $d\vec z_j/dt$ and $\cN^\dag\vec z_j$ as being small.
Then
\begin{eqnarray*}
\cL^\dag\vec z_j^0	 & = & \adot{\cL^\dag\vec z_j^0}{\vec e_k^0}\vec
z_k^0  \\
	 & = & \adot{\vec z_j^0}{\cL\vec e_k^0}\vec z_k^0
	 \quad\mbox{by adjoint property}  \\
	 & = & \adot{\vec z_j^0}{\vec e_\ell^0\cG_{\ell k}}\vec z_k^0
	 \quad\mbox{by (\ref{Egbas})}  \\
	 & = & \delta_{j\ell}\cG_{\ell k}\vec z_k^0
	 \quad\mbox{by orthonormality~(\ref{Ewdef})}
\end{eqnarray*}
Thus the initial approximation for the projection vectors must satisfy
\begin{equation}
	\cL^\dag\vec z_j^0 = \cG_{j k}\vec z_k^0\,.
	\label{Ez0gen}
\end{equation}

For the nonlinear description, define the ``amplitudes'' to be precisely
\begin{equation}
	x=\adot{\vec z_x^0}{\vec u}\,,\
	y=\adot{\vec z_y^0}{\vec u}\,,
	\quad\mbox{where}\quad
	\vec z_x^0=\left[
	\begin{array}{c}
		1/2  \\
		0  \\
		0
	\end{array}
	\right]\,,\quad
	\vec z_y^0=\left[
	\begin{array}{c}
		1/2  \\
		1/2  \\
		0
	\end{array}
	\right]\,,
	\label{Eleadz}
\end{equation}
in terms of original variables $\vec u$.  Using the first part of the
computer algebra program listed in \S\ref{SShopro}, a quadratic
approximation to the nonlinear shape of the centre manifold is
\begin{equation}
	\vec u=\left[
	\begin{array}{l}
		2x   \\
		-2x + 2y  \\
		-3x +  y +\frac{42}{5}y^2+\frac{22}{5}xy+\frac{88}{5}x^2
		+\epsilon y+\epsilon x
	\end{array}
	\right]+\Ord{\epsilon^3+x^3+y^3}\,.
	\label{Ehocm2}
\end{equation}
The lowest order structurally stable model on the centre manifold is
the following cubic model
\begin{equation}
\begin{array}{rcl}
	\dot x & = & -y  - 2 x y + 6 x^2 + \epsilon x
\\&&
-\frac{84}{5}xy^2-\frac{44}{5}x^2y-\frac{176}{5}x^3-2\epsilon xy-2\epsilon x^2
            +\Ord{\epsilon^2+x^4+y^4}\,,  \\
	\dot y & = & x +\epsilon x
	             +\Ord{\epsilon^2+x^4+y^4}\,.
\end{array}
	\label{Ehomod2}
\end{equation}
As is generally the case in Hopf bifurcations, with these correct
cubic nonlinearities this model is usefully predictive.  Numerical
simulations show the birth of a limit cycle as $\epsilon$ crosses
through zero.  Here we have systematically reduced the dynamics by a
small step down from 3-D to a 2-D model.  In serious applications to
very high dimensional dynamics, we would have reduced enormously the
dimensionality of the dynamics.

\subsection{Projection onto oscillating dynamics}
\label{SShopro}

In this subsection I first generalise the iterative algorithm
presented in \S\ref{SSitp} to the case of oscillatory dynamics on the
centre manifold.  To illustrate the algorithm and the typical results I
then apply it to the simple dynamical system~(\ref{Ehopfewig}).

Restart the generic nonlinear analysis from~(\ref{Edzl}).  The
right-hand side is just the residual of the projected dual
equation~(\ref{Ezeq}); it stays the same.  The operator on the
left-hand side has to be simplified, but not as drastically as in
\S\ref{SSitp}.
\begin{itemize}
	\item The third term $\adot{\cD\pz_j}{\vec e_k}\tz_k$ only changes
	that part of the left-hand side in the space spanned by
	$\{\tz_k\}$.  But we do not solve the dual~(\ref{Eadj}) in this
	space, hence the projection seen in~(\ref{Ezeq}).  Thus this term
	is safely ignored.

	\item  The inner product in the second term of~(\ref{Edzl})
	simplifies under approximation as follows.
	\begin{eqnarray*}
		\adot{\cD\tz_j}{\vec e_k}
		& \approx & \adot{\cD\tz_j}{\vec e_k^0}
		\quad\mbox{as $\vec e_k\approx \vec e_k^0$}  \\
		 & \approx & \adot{\cL^\dag\vec z_j^0}{\vec e_k^0}
		 \quad\mbox{neglecting small terms}  \\
		 & = & \adot{\cG_{j\ell}\vec z_{\ell}^0}{\vec e_k^0}
		 \quad\mbox{using (\ref{Ez0gen})} \\
		 & = & \cG_{jk}
		 \quad\mbox{as $\adot{\vec z_{\ell}^0}{\vec
e_k^0}=\delta_{\ell
		 k}$.}
	\end{eqnarray*}

	\item  The first term, the dual operator, approximates to the
	homological operator
	\begin{displaymath}
		\cD\pz_j=(\cG_{k\ell}s_\ell+g_k)\D{\pz_j}{s_k}+\cJ^\dag\pz_j
		\approx \cG_{k\ell}s_\ell \D{\pz_j}{s_k}+\cL^\dag\pz_j\,.
	\end{displaymath}
\end{itemize}
Thus in general we solve
\begin{equation}
	-\cG_{k\ell}s_\ell \D{\pz_j}{s_k}-\cL^\dag\pz_j
	+\cG_{jk}\pz_k
	= \cD\tz_j-\adot{\cD\tz_j}{\vec e_k}\tz_k\,,
	\label{Egendzl}
\end{equation}
for the corrections to the projection vectors.

The complicating feature of~(\ref{Egendzl}) is that it is a coupled
set of equations for the correction vectors $\pz_j$---coupled through
$\cG_{jk}\pz_k$---\emph{and} that it is to be solved in the space of
multinomials in the amplitudes $\vs$---because of the structure of
the homological operator.  As noted earlier, in the iteration we may
ignore components in $\vec z_j^0$, but we must enforce
orthogonality via the iterative correction~(\ref{Epz}).  Other than
these complications the outline of the iterative algorithm is the same
as in the previous section.

For the specific dynamical system~(\ref{Ehopfewig}) a \reduce{}
computer algebra program follows.  Observe that the first part of the
program computes the description of the centre manifold model.  It is
the second part that determines the projection.

{\footnotesize
\verbatimlisting{honew.red}
}

Observe how the second part of this program implements the algorithm,
in lines 50--81.
\begin{enumerate}
\item  $\ell$14--48 compute the centre manifold~(\ref{Ehocm2}) and the
model evolution~(\ref{Ehomod2}) using the iterative algorithm of
\cite{Roberts96a}.  Note that \reduce{} does not implement matrices
well and so $\ell$7--12 define two procedures to help.

\item  $\ell$51--65 is initialisation.
\begin{itemize}
	\item  $\ell$52--53 computes the adjoint operator $\cJ^\dag$ on
$\cM_c$.

	\item  $\ell$54 sets the linear projection vectors $\vec z_j^0$ into
	a $3\times 2$ matrix.

	\item  $\ell$55--57 computes the tangent vectors $\vec e_j$ of $\cM_c$.

	\item  $\ell$59 says to truncate the computations to have errors
	$\Ord{\delta^3}$ as these should be accurate enough and will speed
	computation.

	\item  $\ell$60--63 define general multinomial expressions and
	a list of their coefficients for later use in solving the homological
	equation via the method of undetermined coefficients.

	\item $\ell$65 sets $\vec z_j$ to its initial approximation.
\end{itemize}

\item $\ell$66--81 perform the iterations until the residuals of the
desired equations are zero to the order of error specified.
\begin{enumerate}
\item $\ell$68--70 computes the residuals of the projected dual
equation~(\ref{Ezeq}) and the orthonormality condition~(\ref{Ewdef}),
assigned to \verb|rdz| and \verb|rze| respectively.

	\item  $\ell$72--77 solves for the undetermined coefficients of the
	corrections using the third component of equation~(\ref{Egendzl}).
	The first term on the right-hand side of $\ell$70, for example,
	comes from $-\cL^\dag\pz_j$; the second and third terms represent
	$-\cG_{k\ell}s_\ell \D{\pz_j}{s_k}$; while the fourth term comes
	from $\cG_{jk}\pz_k$; and the last term is from the residual.

	\item  $\ell$79--80 updates the approximation to better satisfy the
	orthonormality and the projected dual.
\end{enumerate}
\end{enumerate}

The output of this computer program indicates that projections from
initial conditions off $\cM_c$ onto $\cM_c$ are to be orthogonal to
\begin{equation}
	\vec z_x\approx\left[
	\begin{array}{cll}
		\frac{1}{2}
		&+\frac{2}{5}y-\frac{6}{5}x

	&-\frac{26}{25}y^2-4xy+\frac{714}{25}x^2-\frac{28}{25}\epsilon y
		+\frac{34}{25}\epsilon x\\[4pt]
		0
		&-\frac{1}{5}y+\frac{3}{5}x
		& -\frac{124}{25}y^2+\frac{68}{5}xy+\frac{6}{25}x^2
		+\frac{4}{25}\epsilon y+\frac{13}{25}\epsilon x\\[4pt]
		0
		&+\frac{2}{5}y-\frac{6}{5}x
		&+\frac{16}{5}y^2-\frac{44}{5}xy+\frac{24}{5}x^2
		-\frac{18}{25}\epsilon y+\frac{4}{25}\epsilon x
	\end{array}
	\right]\,,
	\label{Ezxeg}
\end{equation}
\begin{equation}
	\vec z_y\approx\left[
	\begin{array}{cll}
		\frac{1}{2}
		&-\frac{4}{5}y+\frac{2}{5}x
		&+\frac{52}{25}y^2+\frac{74}{5}xy-\frac{238}{25}x^2
		+\frac{16}{25}\epsilon y+\frac{2}{25}\epsilon x\\[4pt]
		\frac12
		&+\frac{2}{5}y-\frac{1}{5}x
		&+\frac{248}{25}y^2-\frac{16}{5}xy-\frac{2}{25}x^2
		+\frac{12}{25}\epsilon y-\frac{11}{25}\epsilon x\\[4pt]
		0
		&-\frac{4}{5}y+\frac{2}{5}x
		&-\frac{32}{5}y^2+\frac{16}{5}xy-\frac{8}{5}x^2
		-\frac{4}{25}\epsilon y+\frac{12}{25}\epsilon x
	\end{array}
	\right]\,.
	\label{Ezyeg}
\end{equation}

\begin{figure}[tbp]
	\centering
	\includegraphics{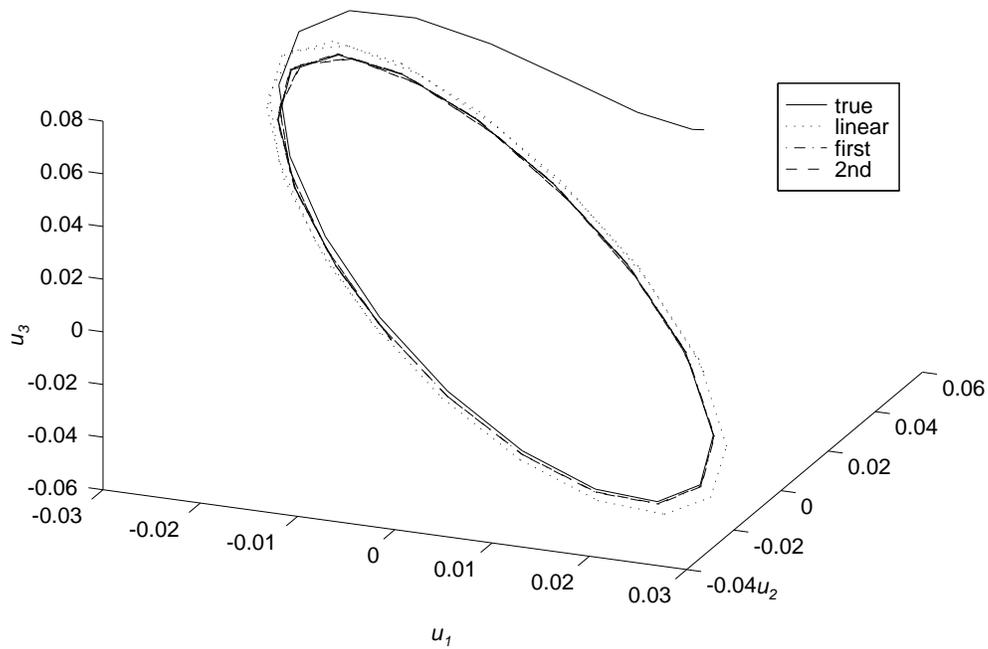} \caption{trajectories
	of~(\ref{Ehopfewig}) from the ``true'' initial condition of $\vec
	u_0$ (solid), and from the various projections onto $\cM_c$:
	linear from $\vec z^0_j$ (dotted); first order (dot-dashed);
	second order (dashed).  Observe the approximately exponential
	approach between the ``true'' and the model trajectories, but that
	the linearly projected trajectory is slightly awry.}
	\label{Fcftraj}
\end{figure}

\begin{figure}[tbp]
	\centering
	\includegraphics{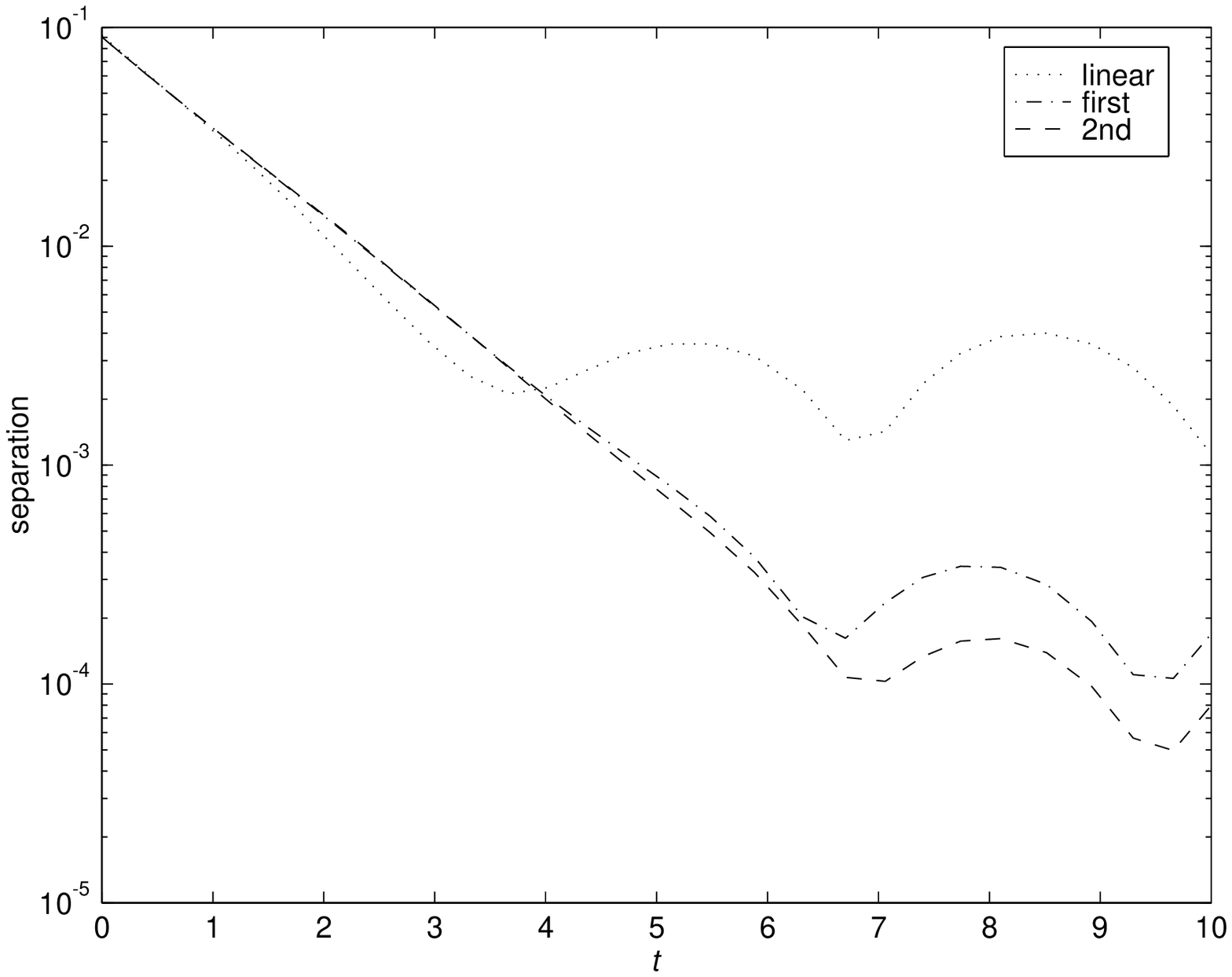} \caption{log of the separation
	between the trajectories shown in Figure~\ref{Fcftraj} and the
	``true'' solution: linear from $\vec z^0_j$ (dotted); first order
	(dot-dashed); second order (dashed).  Observe the initial
	exponential approach, but that the first order and second order
	approximations to the initial conditions of the model are an order
	of magnitude better.}
	\label{Fseptraj}
\end{figure}

For example, consider the dynamics from the initial condition $\vec
u_0=(0.022,0,0.073)$ and the corresponding evolution on $\cM_c$.
According to either the leading order projection or the definition of the
amplitudes $x$ and $y$, equation~(\ref{Eleadz}), $\vec u_0$
corresponds to the point on $\cM_c$ with parameters $x=y=0.011$,
namely $\vec u_0^0=(0.022,0,-0.019)$.  However, at this point on
$\cM_c$ the projection of initial conditions should be slightly
different; according to the linear modifications
in~(\ref{Ezxeg}--\ref{Ezyeg}) it should be orthogonal to the columns
of
\begin{displaymath}
	\vec z=\left[
	\begin{array}{rr}
		0.492 & 0.496  \\
		0.004 & 0.502  \\
		-0.008 & 0.004
	\end{array}
	\right]\,.
\end{displaymath}
Thus, according to~(\ref{Esolv}), a more appropriate initial condition
on $\cM_c$ is $\vec u_0^1=(0.021,0.001,-0.017)$.  Similarly, the
second order corrections in~(\ref{Ezxeg}--\ref{Ezyeg}) refine the initial
conditions further.  The differences here are rather small, but the
improvement is seen in Figures~\ref{Fcftraj}--\ref{Fseptraj}.
Figure~\ref{Fcftraj} shows the qualitative picture of the trajectory
starting at $\vec u_0$ exponentially quickly approaching $\cM_c$, but
that the model is only quantitatively predictive if $\vec u_0$ is
projected correctly.  Figure~\ref{Fseptraj} quantifies a comparison
between the various orders of approximation to the projection and
demonstrates that the refined projection vectors $\vec z_j$ do
perform better.

Also recall from \S\ref{SSforc} that even such small corrections to
the projection as we see here may have a considerably larger influence
when projecting an applied forcing onto the centre manifold model.

\section{Concluding remarks}
\label{Sconc}

Work in progress will show how to apply the techniques presented here
to problems of more physical interest.  In particular I am examining
the issue of projecting initial conditions onto the lubrication
model~\cite{Roy96} of the flow of a thin film of fluid over a solid
substrate.  Although there are no inertia effects in the lubrication
model of the dynamics, there may well be such effects in the provision
of initial conditions as the fluid dynamics relaxes to lubrication
flow.  Additionally, the effects of substrate roughness on the flow
may be determined by projecting the appropriately perturbed substrate
boundary conditions.

Finding and coding the adjoint $\cJ^\dag$ can be a major headache,
especially for problems such as free-surface fluid dynamics.  I have
sought approaches based directly upon the Jacobian $\cJ$ rather than
its adjoint, because then the residual driving the iteration could be
determined directly and very simply from the governing equations.
However, so far, the only method I have found also involves
determining the equivalent of $\vec w$, introduced in~(\ref{Efmod}),
which is considerably more involved.

Throughout this paper we have addressed the projection of
initial conditions and forcing, \emph{linearly} correct in distance
from the centre manifold or in forcing amplitude.  If one needs
to determine effects \emph{nonlinear} in distance, then a more
sophisticated analysis is needed.  At this stage the only approach I
can imagine also involves the considerable labour of finding $\vec w$.

Lastly, Muncaster and Cohen \cite{Muncaster83c,Cohen88} suggested the
construction of the low-di\-men\-sion\-al manifold of slow, rigid-body
dynamics by neglecting rapidly oscillating modes.  An extremely simple
example of the motion of a one-di\-men\-sion\-al elastic body is
discussed in \cite[\S2]{Roberts93}.  In contrast to the rapid collapse
to the centre manifold, the slow dynamics on the slow manifold form a
low-dimensional model because they act as a ``centre'' for the fast
oscillations of neighbouring trajectories.  This principle of
neglecting fast oscillations is precisely equivalent to the guiding
centre principle of Van Kampen \cite{vanKampen85}.  The algebra needed
to develop slow manifold models is identical to that presented here
and in \cite{Roberts96a}.  The algebra to model initial conditions
will also be the same.  However, slow manifolds are much more
delicate.  Cox \& I \cite{Cox93a,Cox93b} used normal forms to show
that the dynamics on and off the slow manifold generally differ by an
amount of $\Ord{\varepsilon^2}$, where $\varepsilon$ measures the
amplitude of the fast oscillations, it measures the distance off
$\cM_0$.  That is, generically there is some unavoidable slip between
a slow model and the fully detailed oscillating dynamics.

\bibliographystyle{plain}\bibliography{ajr,bib,new}

\end{document}